\documentclass[onecolumn]{svjour3}
\usepackage{amssymb,amsmath}

\begin{document}

\title{Higher-order theories from the minimal length}
\author{M. Dias$^1$ \and J. M. Hoff da Silva$^2$ \and E. Scatena$^3$}
\institute{$^1$Departamento de Ci\^encias Exatas e da Terra, Universidade Federal de S\~ao Paulo\\
Diadema-SP-Brazil.\\
$^2$Departamento de F\'isica e Qu\'imica, Universidade Estadual Paulista ``J\'ulio de Mesquita Filho'' - UNESP, Guaratinguet\'a, SP - Brazil\\
$^3$Universidade Federal de Santa Catarina, Blumenau - SC - Brazil\\
\email{marco.dias@unifesp.br} \and \email{hoff@feg.unesp.br}\and \email{e.scatena@ufsc.br}}
\date{\today}
\maketitle

\begin{abstract}
We show that the introduction of a minimal length in the context of non-commutative spacetime gives rise (after some considerations) to higher-order theories. We then explicitly demonstrate how these higher-derivative
theories appear as a generalization of the standard electromagnetism and general relativity by
applying a consistent procedure that modifies the original Maxwell and Einstein-Hilbert actions. In order to set a bound on the minimal length, we compare the deviations from the inverse-square law with the potentials obtained in the higher-order theories and discuss the validity of the results. The introduction of a quantum bound for the minimal length parameter $\sqrt{\beta}$ in the higher-order QED allows us to lower the actual limits on the parameters of higher-derivative gravity by almost half of their order of magnitude.
\end{abstract}


\section{Introduction}

Higher-order theories may arise from several different considerations but, in essence, they can be cast in two major categories: an \emph{ad hoc} procedure made to parametrize small deviations from a lower order theory, or truncated series from a perturbative expansion of a given effective theory. Although in a classical level these higher-order terms can be faced as small corrections to the original theory, in the quantum realm their behaviour introduce some serious (and undesirable) consequences such as non-unitarity and non-locality \cite{simon}.

Despite those problems, higher-order theories have a crucial advantage: its potential is finite on the origin, meaning that we do not have a divergence upon the charge or mass. From the quantum point of view, it renders the theory directly power-counting renormalizable (i.e., its propagator behaves like $1/k^4$). Moreover, one can speculate that this finiteness must be related to the properties of the spacetime itself at small distances, near the charge, forbidding the theory from explode on the origin. 

On the other hand, those features related to higher-derivative theories also appear in the scope of non-commutative theories, and questions concerning unitarity and causality have been widely discussed in this context \cite{gaume,gomis,chu}. However, in dealing with non-commutative spacetime, we already expect that usual notions of locality and causality do not hold, since they manifest themselves at the Planck scale, where the very concept of geometry must be reviewed.

Therefore, it should not be astonishing (although unexpected), that the generalized commutation relation between $X^{\mu}$ and $P^{\mu}$ (appearing via the introduction of a minimal length parametrized by $\beta$) leads to higher-order theories, with all of its idiosyncrasies, but from considerations entirely different from the usual ones. Consequently, higher-order theories naturally emerge from the non-commutative scenario,   and are not ``placed by hand'', contrary to usual proposed models \cite{grinstein}.

Going further, it is known that Lorentz symmetry has been confirmed in all energy scales accessible nowadays, despite the effort that have been made to detect violations of it \cite{nature,kostelecky}. That been said, if one is willing to abandon the assumption of a continuous spacetime, it is possible to construct a Lorentz-covariant quantized spacetime, parametrized by a minimal length $\sqrt{\beta}$, in which the coordinates do not commute, giving rise to a non-commutative theory \cite{snyder,Quesne:2006fs}.

In this vein, we show in Section II that the generalized commutation relation between spacetime points and momentum originate a new derivative operator. Interestingly enough, to the first order in $\beta$ we can guarantee that the position operators do commute, and in the space representation the new derivative operator contains, in addition to the ordinary derivative, a D'Alembertian weighted by the parameter $\beta$ and the Planck's constant $\hbar$. 

Therefore, it is possible to see how the non-commutativity manifests itself, at least at first order in $\beta$, in the usual physical theories upon the replacement of the usual derivatives by the generalized one. That means that even without construct  a non-commutative field theory explicitly, we still can grasp its effects and, moreover, set bounds on the minimal length considering the higher-derivative theories that it generates.

In Section III we proceed with the aforementioned replacement of derivatives for the Einstein-Hilbert's action in (3+1)D, rendering a higher-order theory of gravity. Using the experimental data available for the validity of the Newtonian potential in Cavendish-like experiments with torsion balances, we can set a gravitational bound on the minimal length. We also can relate the bounds on the higher-derivative gravity parameters (obtained from the deflection of light by the Sun and the gravitational Doppler effect) to set a limit on the minimal length.

The same procedure is done in Section IV for the Maxwell's action, and it is shown that it gives rise to a higher-order theory of electromagnetism (e.g., Lee-Wick or Podolsky electrodynamics). Since the experimental results for the validity of the Coloumb's law are more precise than those for the Newtonian potential we find a better bound on $\sqrt{\beta}$. Besides, a higher-derivative version of Quantum Electrodynamics allows us to set a more stringent limit on the minimal length, due to the excellent precision of the anomalous magnetic moment of the electron.

Finally, in Section V, we summarize the limits obtained for the minimal length. As a spin-off, taking advantage of QED results, we use the quantum bound for $\beta$ to set new constraints on the usual higher-order gravity parameters $\alpha R^2$ and $\gamma R_{\mu\nu}^2$, and discuss the validity of the obtained results.

\section{Non-commutative relations in the position space}

It is well known that the generalization of the Lorentz algebra encompassing a minimal length leads to a non-commutative generalized relation between $X^{\mu}$ and $P^{\mu}$ \cite{Quesne:2006fs}. In fact, in the momentum representation $(P^{\mu}=p^{\mu})$ it is possible to show that the commutation relations are given by 
\begin{equation}
[X^{\mu},P^{\nu}]=-i\hbar [(1-\beta p_{\rho}p^{\rho})\eta^{\mu\nu}-2\beta p^{\mu}p^{\nu}],\label{j1}
\end{equation} where the parameter $\beta$ encodes the minimal length. The other relations are given by
\begin{equation}
[X^{\mu},X^{\nu}]=i\hbar \frac{2\beta-\beta'-(2\beta+\beta')\beta p_{\rho}p^{\rho}}{1-\beta p_{\rho}p^{\rho}}(p^{\mu}x^{\nu}-p^{\nu}x^{\mu}),\label{j2} 
\end{equation} and, obviously, $[p^{\mu},p^{\nu}]=0$. In this case, the $X^{\mu}$ operator is given by \begin{equation}
X^{\mu}=(1-\beta p_{\nu}p^{\nu})x^{\mu}-\beta'p^{\mu} p_{\nu}x^{\nu}+i\hbar \gamma p^{\mu}.\label{j3} 
\end{equation} 

There are two important remarks on the Eqs. (\ref{j1}-\ref{j2}). Firstly, there is a specific point in the parameter space, namely $\beta'=2\beta$, in which the generalized position operators commute up to a $\beta^{2}$ term, i.e. $[X^{\mu},X^{\nu}]=0+O(\beta^{2})$. Besides, the $\gamma$ parameter appearing in Eq. (\ref{j3}) does not intervene in the commutation relations. Both these aspects will be important in what follows. 

In order to obtain the functional form of the generalized momentum operator in the space representation $(X^{\mu}=x^{\mu})$ we start from the following Ansatz \begin{equation} P^{\mu}=(1-\beta p_{\alpha}p^{\alpha})Z^{\mu}-\beta'K^{\mu}p_{\alpha}x^{\alpha}+i\hbar \gamma Q^{\mu}.\label{j4} \end{equation} The reason for this specific ansatz is simple. To begin with, it respects the same tensorial structure of Eq. (\ref{j3}), already taking into account the current contractions. Moreover, the unknown quantities are to be discovered by the requirement that the commutation relation (\ref{j1}) be recovered. Hence, after a bit of usual algebra one has 
\begin{eqnarray}
[X^{\mu},P^{\nu}]&=&\left.[x^{\mu},p^{\nu}]-\beta (x^{\mu}p_{\alpha}p^{\alpha}Z^{\nu}-p_{\alpha}p^{\alpha}Z^{\nu}x^{\mu})\right.\nonumber\\&-&\left.\beta'(x^{\mu}K^{\nu}p_{\alpha}x^{\alpha}-K^{\nu}p_{\alpha}x^{\alpha}x^{\mu})\right.\nonumber\\&+&\left.i\hbar \gamma [x^{\mu},Q^{\nu}].\right. \label{j5}
\end{eqnarray}  Comparing this last equation with Eq. (\ref{j1}) we can see that a possible choice for $Q^\nu$ is $Q^\nu \sim x^\nu$, since $\gamma$ does not enter in the commutation relation. Hence, let us write $Q^\nu=qx^\nu$, where $q$ is an arbitrary constant. Now, we shall handle the term $x^\mu p_\alpha p^\alpha Z^\nu$ appearing in the right hand side (RHS) of Eq. (\ref{j5}). Using a very standard trick it reads 
\begin{eqnarray}
x^\mu p_\alpha p^\alpha Z^\nu&=&([x^{\mu},p_{\alpha}]+p_{\alpha}x^\mu)p^\alpha Z^\nu \nonumber
\\ &=&[x^{\mu},p_{\alpha}]p^\alpha Z^\nu+p_\alpha ([x^\mu,p^\alpha]+p^\alpha x^\mu)Z^\nu,\nonumber
\end{eqnarray} which can finally be recast into the following way 
\begin{equation}
x^\mu p_\alpha p^\alpha Z^\nu=-2i\hbar p^\mu Z^\nu+p_\alpha p^\alpha x^\mu Z^\nu .\label{j6}
\end{equation}
Hence, the commutation relation reads \begin{eqnarray} [X^\mu,P^\nu]&=&\left.(1-\beta p_{\alpha}p^\alpha)[x^\mu,Z^\nu]+2\beta i \hbar p^\mu Z^\nu\right.\nonumber\\&-&\left.2\beta(x^\mu K^\nu p_\alpha x^\alpha-K^\nu p_\alpha x^\alpha x^\mu).\label{j7} \right. \end{eqnarray} 
In view of Eq. (\ref{j7}), by comparing it with (\ref{j1}), we conclude that $Z^\mu=p^\mu$ and $K^\mu=0$. Moreover, in order to avoid complications from the commutation relation of $P$'s operators, we shall particularize our analysis to the $\beta'=2\beta$ case. So, after all, we arrive at the functional form of the $P$ operator, given by 
\begin{equation}
P^\mu = (1-\beta p_\alpha p^\alpha)p^\mu +i\hbar \gamma q x^\mu,\label{j8} \end{equation} 
or, by means of its quantum operator 
\begin{equation} 
\nabla_\mu=(1+\beta\hbar^2\Box)\partial_\mu -\gamma q x_\mu.\label{j9}
\end{equation} 
Now, let us call attention for a technicality. We shall apply the Eq. (\ref{j9}) as the generalized derivative operator in usual Lagrangian, in order to see whether the minimal length encoded in (\ref{j9}) leads to higher derivative models. In order to accomplish this program, we shall force $q=0$, otherwise we are not able to reproduce a gauge invariant theory. Note that, even taking $Q_\mu=x_\mu$ (without any reference to the constant $q$) it is possible to get rid of the last term of (\ref{j9}) taking $\gamma=0$. Notice that in doing so (taking $\gamma=0$) one is not violating any requirement of the generalized Lorentz algebra. In particular, the demand of hermiticity of $X$ and $P$ operators with respect to a new weighed inner product is not spoiled by the choice $\gamma =0$. Therefore, the new derivative may be written as 
\begin{equation} 
\nabla_\mu=(1+\beta\hbar^2\Box)\partial_\mu,\label{j10} 
\end{equation} i.e. a direct generalization of the one used in Ref. \cite{Moayedi:2012fu}. We reinforce that in order to arrive at the prescription (\ref{j10}) use was made of the specific relation $\beta'=2\beta$ and $\gamma=0$. Therefore, after all, we are dealing with a possible but restrictive case in the parameter space. Also, in a broader context, it seems quite plausible to recover (\ref{j10}) from different considerations.

In what follows, we proceed to apply the generalized derivative $\nabla_{\mu}$ from eq. (\ref{j10}) in two distinct cases: gravitation and electromagnetism.

\section{The Gravitational Case}

The application of the generalized derivative operator encoding the minimal length to the gravitational case is more subtle. The reason rest upon the fact that in the linearized theory, in which we shall investigate the graviton propagator, it is not clear, at first sight, whether one need to linearize before implementing the generalized derivative, or after that. As we shall see, however, the final result is independent of the chosen order.

We start implementing the minimal length derivative in the linearized version of the Einstein-Hilbert $(3+1)-$dimensional Lagrangian. After that, we demonstrate that the result is the same one of taking the linearization of the Einstein-Hilbert plus generalized derivative action. It is worth to note that in the linearized theory of gravitation we are essentially treating a spin-2 field in a flat Minkowski space and, thus the substitution of the ordinary derivative by the generalized one, which is Lorentz covariant by construction, should proceed without any inconsistency. Therefore, it should be clear that in this weak field approximation it is not necessary to treat the non-commutative curved space theory, a subject with interesting subtleties on its own.

\subsection{Procedure one: starting from the linearized Lagrangian}

In this case, after expanding the metric in the form
\begin{equation}
g_{\mu\nu}=\eta_{\mu\nu}+\kappa h_{\mu\nu},\label{j16}
\end{equation} where $\kappa$ is related to the gravitational constant ($\kappa^2=32\pi G$), it can be easily verified that 
\begin{equation}
g^{\mu\nu}=\eta^{\mu\nu}-\kappa h^{\mu\nu}+\kappa^2h^{\mu\alpha}h_{\alpha}^{\;\; \nu}+\cdots,\label{j17}
\end{equation}
\begin{equation}
\Gamma_{\mu\alpha}^{\beta}=\frac{\kappa}{2}[h_{\mu\;\;,\alpha}^{\;\;\beta}+h_{\alpha\;\;,\mu}^{\;\;\beta}-h_{\mu\alpha}^{\;\;\;\;\;,\beta}],\label{j18}
\end{equation}
\begin{equation}
\sqrt{-g}=1+\frac{\kappa}{2}h-\frac{\kappa^2}{4}h^{\alpha\beta}h_{\alpha\beta}+\frac{\kappa^2}{8}h^2+\cdots .\label{j19}
\end{equation}
Now, starting from the gamma-gamma version of the Einstein-Hilbert action 
\begin{equation}
S=\frac{2}{\kappa^2}\int d^4x \sqrt{-g} g^{\mu\nu} \Big[-\Gamma_{\mu\alpha}^{\beta}\Gamma_{\beta\nu}^{\alpha}+\Gamma_{\mu\nu}^{\beta}\Gamma_
{\beta\alpha}^\alpha\Big],\label{j20}
\end{equation} we work with its linearized Lagrangian version given by 
\begin{eqnarray}
\mathcal{L}_{lin}&=&\left.\frac{1}{2}\Big[-\Big(h_{\beta}^{\;\;\alpha,\mu}+h^{\mu\alpha}_{\;\;\;\;\;,\beta}-h_{\beta}^{\;\;\mu,\alpha}\Big)\Big(h_{\mu\;\;,\alpha}^{\;\;\beta}+h_{\alpha\;\;,\mu}^{\;\;\beta}-h_{\mu\alpha}^{\;\;\;\;,\beta}\Big)\right. \nonumber\\&+&\left. h_{,\beta}\Big(h_{\mu}^{\;\;\beta,\mu}+h^{\mu\beta}_{\;\;\;\;,\mu}-h^{,\beta}\Big)\Big]\right. .\label{j21} 
\end{eqnarray} Implementing the generalized derivative means $h_{\beta}^{\;\;\alpha,\mu}\rightarrow (1+\beta \hbar^2\Box)h_{\beta}^{\;\;\alpha,\mu}$, the net effect in the above Lagrangian reads 
\begin{eqnarray}
\nonumber \mathcal{L}&=&\frac{1}{2}\left[-(1+\beta\hbar^2\Box)\Big(h_{\beta}^{\;\;\alpha,\mu}+h^{\mu\alpha}_{\;\;\;\;\;,\beta}-h_{\beta}^{\;\;\mu,\alpha}\Big)(1+\beta\hbar^2\Box)\Big(h_{\mu\;\;,\alpha}^{\;\;\beta}\right. + \\ &+& \left. h_{\alpha\;\;,\mu}^{\;\;\beta}-h_{\mu\alpha}^{\;\;\;\;,\beta}\Big)+(1+\beta\hbar^2\Box)h_{,\beta}(1+\beta\hbar^2\Box)\Big(h_{\mu}^{\;\;\beta,\mu}+h^{\mu\beta}_{\;\;\;\;,\mu}-h^{,\beta}\Big)\right]. \label{j22} 
\end{eqnarray} Separating out the standard terms we have, disregarding $O(\beta^2\hbar^4)$ terms, 

\begin{eqnarray}
\nonumber\mathcal{L}&=&\left.\mathcal{L}_{lin}+\frac{1}{2}\Bigg\{-\beta\hbar^2\Big(h_{\beta}^{\;\;\alpha,\mu}+h^{\mu\alpha}_{\;\;\;\;\;,\beta}-h_{\beta}^{\;\;\mu,\alpha}\Big)\Box\Big(h_{\mu\;\;,\alpha}^{\;\;\beta}+h_{\alpha\;\;,\mu}^{\;\;\beta}-h_{\mu\alpha}^{\;\;\;\;,\beta}\Big)\right.+\\
\nonumber &-&\beta\hbar^2\Box\Big(h_{\beta}^{\;\;\alpha,\mu}+h^{\mu\alpha}_{\;\;\;\;\;,\beta}-h_{\beta}^{\;\;\mu,\alpha}\Big)\Big(h_{\mu\;\;,\alpha}^{\;\;\beta}+h_{\alpha\;\;,\mu}^{\;\;\beta} h_{\mu\alpha}^{\;\;\;\;,\beta}\Big)+\\ 
&+& \left.\beta\hbar^2h_{,\beta}\Box \Big(h_{\mu}^{\;\;\beta,\mu}+h^{\mu\beta}_{\;\;\;\;,\mu}-h^{,\beta}\Big)+\beta\hbar^2\Box h_{,\beta}\Big(h_{\mu}^{\;\;\beta,\mu}+h^{\mu\beta}_{\;\;\;\;,\mu}-h^{,\beta}\Big)\Bigg\}\right.,\label{j23}
\end{eqnarray} 
where $\mathcal{L}_{lin}$ stands for the linearized Lagrangian given by (\ref{j21}). This expression may be considerable simplified using the fact that $\Box fg+f\Box g=\Box(fg)-2\partial_\gamma f\partial^\gamma g$. Thus, after some algebra it reads 
\begin{eqnarray}
\nonumber\mathcal{L}&=& \mathcal{L}_{lin}+\beta\hbar^2\Box\mathcal{L}_{lin}+\beta\hbar^2\Big\{(h_{\beta\alpha,\mu\gamma}+h_{\mu\alpha,\beta\gamma}-h_{\beta\mu,\alpha\gamma})(h^{\mu\beta,\alpha\gamma}+h^{\alpha\beta,\mu\gamma}-h^{\mu\alpha,\beta\gamma})+\\ &-& h_{,\beta\gamma}(2h_{\mu}^{\;\;\beta,\mu\gamma}-h^{,\beta\gamma})\Big\}, \label{j24}
\end{eqnarray}
where, obviously, the second term of the RHS may be dropped under integration over the spacetime volume. The $\mathcal{L}_{lin}$ term is nothing but the usual (first order derivative) one, reproducing the standard Lagrangian. The remain is the higher derivative contribution arising from the fact that the derivative operator encodes the minimal length, just as in the electromagnetic case. Now, we shall move forward inverting our procedure, i.e., starting from the Einstein-Hilbert action with the generalized derivative and then linearizing the resulting action.

\subsection{Procedure two: starting from the generalized Einstein-Hilbert action} 

In the beginning of this section we shall use the notation $\bar{A}$ for a given quantity $A$ which encodes the generalized derivative operator $\nabla_{\mu}$ (\ref{j10}). The action reads 
\begin{equation}
S=\frac{2}{\kappa^2}\int d^4x\sqrt{-g} g^{\mu\nu}\bar{\bar{R}}_{\mu\nu}, \label{j25} 
\end{equation}
where $\bar{\bar{R}}_{\mu\nu}$ is given by 
\begin{eqnarray}
\bar{\bar{R}}_{\mu\nu}&=&\left.-(1+\beta\hbar^2\Box)\bar{\Gamma}^\alpha_{\mu\nu,\alpha}+(1+\beta\hbar^2\Box)\bar{\Gamma}^\alpha_{\mu\alpha,\nu}-\bar{\Gamma}^\beta_{\mu\nu}\bar{\Gamma}^\alpha_{\beta\alpha}+\bar{\Gamma}^\beta_{\mu\alpha}
\bar{\Gamma}^\alpha_{\beta\nu},\right. \label{j26}
\end{eqnarray}
being the $(1+\beta\hbar^2\Box)$ terms coming from the generalized derivative, in such a way that $\bar{\Gamma}^\alpha_{\mu\nu,\alpha}=\partial_\alpha\bar{\Gamma}^\alpha_{\mu\nu}$. Denoting $H^{\mu\nu}=\sqrt{-g}g^{\mu\nu}$ it can be verified that 
\begin{eqnarray}
H^{\mu\nu}\bar{\bar{R}}_{\mu\nu}\!=\!H^{\mu\nu}\bar{R}_{\mu\nu}\!+\!\beta\hbar^2H^{\mu\nu}\Box
\Big(\!\!-\!\bar{\Gamma}^\alpha_{\mu\nu,\alpha}\!+\!\bar{\Gamma}^\alpha_{\mu\alpha,\nu}\!\Big)\!.\label{j27}
\end{eqnarray} 
It is useful to work out the action terms separately. Notice that, apart from a total derivative, we have 
\begin{eqnarray}
H^{\mu\nu}\bar{R}_{\mu\nu}&=&H^{\mu\nu}\Big(-\bar{\Gamma}^\beta_{\mu\nu}\bar{\Gamma}
^\alpha_{\beta\alpha}+\bar{\Gamma}^\beta_{\mu\alpha}\bar{\Gamma}^\alpha_{\beta\nu}\Big)+ H^{\mu\nu}_{\;\;,\alpha}\bar{\Gamma}^\alpha_{\mu\nu}-H^{\mu\nu}_{\;\;,\nu}\bar{\Gamma}^\alpha_{\mu\alpha} . \label{j28}
\end{eqnarray} 
Now, since the covariant derivative of $H^{\mu\nu}$ vanish (the theory still metric) it is possible to write 
\begin{eqnarray}
H^{\mu\nu}_{\;\;,\alpha}&=&H^{\mu\nu}\bar{\Gamma}^{\beta}_{\beta\alpha}-H^{\mu\beta}\bar{\Gamma}^\nu_{\beta\alpha}-H^{\beta\nu}\bar{\Gamma}^\mu_{\beta\alpha}-\beta\hbar^2\Box H^{\mu\nu}_{\;\;,\alpha}.\label{j29}
\end{eqnarray} 
Hence, taking into account Eqs. (\ref{j27}), (\ref{j28}), and (\ref{j29}) we have the following action
\begin{eqnarray}
\nonumber S&=&\frac{2}{\kappa^2}\int d^4x\left\lbrace H^{\mu\nu}\Big[-\bar{\Gamma}_{\mu\alpha}^{\beta}\bar{\Gamma}_{\beta\nu}^{\alpha}+
\bar{\Gamma}_{\mu\nu}^{\beta}\bar{\Gamma}_{\beta\alpha}^{\alpha}\Big]\!+\!\beta\hbar^2
H^{\mu\nu}\Box \Big(\!-\!\bar{\Gamma}_{\mu\nu,\alpha}^\alpha\!+\!\bar{\Gamma}_{\mu\alpha,\nu}^{\alpha}\Big)\right. +\\ &+& \left. \beta\hbar^2\Box H^{\mu\nu}_{\;\;\;\;,\nu}\bar{\Gamma}_{\mu\alpha}^{\alpha}
\!-\!\beta\hbar^2\Box H^{\mu\nu}_{\;\;\;\;,\alpha}\bar{\Gamma}_{\mu\nu}^\alpha\right\rbrace.\label{j30}
\end{eqnarray} 
Before linearizing the gravitational action it is necessary to make explicit every term with the generalized derivative. In this vein, notice that 
\begin{eqnarray} 
\bar{\Gamma}_{\mu\nu}^{\sigma}=\Gamma_{\mu\nu}^{\sigma}+\frac{\beta\hbar^2}{2}g^{\sigma\lambda}\Box\Big(\partial_{\mu}g_{\lambda\nu}+\partial_\nu g_{\lambda\mu}-\partial_\lambda g_{\mu\nu}\Big).\label{j31}
\end{eqnarray}
Therefore it is possible to see that the $\bar{\Gamma}\bar{\Gamma}$ terms of Eq. (\ref{j30}) will give rise to the action (\ref{j20}), which we shall call $S_{GG}$. The entire resulting action, up to $\beta\hbar^2$ terms, is given by 
\begin{eqnarray} 
\nonumber S&=&S_{GG}+\frac{\beta\hbar^2}{\kappa^2}\int d^4x H^{\mu\nu}\left\lbrace -\Gamma_{\mu\alpha}^{\beta}g^{\alpha\gamma}\Box(\partial_{\beta}g_{\gamma\nu}+\partial_{\nu}
g_{\gamma\beta}-\partial_{\gamma}g_{\nu\beta})\right. + \\ 
\nonumber &-& \Gamma_{\beta\nu}^\alpha g^{\beta\gamma}\Box(\partial_{\mu}g_{\gamma\alpha}+\partial_{\alpha}g_{\gamma\mu}-
\partial_{\gamma}g_{\mu\alpha})+\Gamma_{\mu\nu}^{\beta}g^{\alpha\gamma}\Box(\partial_{\beta}g_{\gamma\alpha}+\partial_{\alpha}
g_{\gamma\beta}-\partial_{\gamma}g_{\alpha\beta}) + \\ 
\nonumber &+&  \left. \Gamma_{\beta\alpha}^\alpha g^{\beta\lambda}\Box(\partial_{\mu}g_{\lambda\nu}+\partial_\nu g_{\lambda\mu}-\partial_\lambda g_{\mu\nu})\right\rbrace +\frac{2\beta\hbar^2}{\kappa^2}\int d^4x \left\lbrace H^{\mu\nu}\Box\Gamma^\alpha_{\mu\alpha,\nu} \right. +\\ &+& \left.\Box H^{\mu\nu}_{\;\;\;\;,\nu}\Gamma^\alpha_{\mu\alpha}-H^{\mu\nu}\Box\Gamma^\alpha_{\mu\nu,\alpha}-\Box H^{\mu\nu}_{\;\;\;\;,\alpha}\Gamma^\alpha _{\mu\nu}\right\rbrace . \label{j32}
\end{eqnarray}
Now we are in position to analyse the linearization by means of Eqs. (\ref{j16}), (\ref{j17}), (\ref{j18}) and (\ref{j19}). The procedure is quite standard and the calculations are lengthy. The interesting point to be remarked, however, is that the last four terms of the RHS of Eq. (\ref{j32}) cancel out each other and, therefore, do not contribute to the linearized version of the action. Thus, after some manipulations, we are left simply with 
\begin{eqnarray}
\nonumber S_{lin}&=&S_{GGlin}+\beta\hbar^2\int d^4x\left\lbrace\Big(h^{\gamma\alpha,\mu\lambda}+h^{\mu\alpha,\gamma\lambda}-h^{\gamma\mu,\alpha\lambda}\Big)\Big(h_{\gamma\alpha,\mu\lambda}+h_{\gamma\mu,\alpha\lambda}-h_{\mu\alpha,\gamma\lambda}\Big)\right. + \\  &-& \left. h_\mu^{\;\;\beta,\mu\gamma}h_{,\beta\gamma}-\Big(h_{\mu\beta}^{\;\;\,\;\;\,,\mu}-h_{,\beta}\Big)_{\!,\gamma}\,h^{,\beta\gamma} \right\rbrace , \label{j33}
\end{eqnarray}
which is, obviously, the action for the Lagrangian (\ref{j24}). We reinforce that, interestingly enough, the higher derivative terms are included \emph{ab initio} in the lagrangian, making possible its use in the propagator procedure in a usual fashion.

Once the final form of the Lagrangian is found, we can immediately see that it has the same form of other proposed models of higher-order gravity \cite{Stelle}. Obviously, the properties of the higher-order model that we have found here will not be entirely different from the usual ones, and questions concerning features like propagating massive ghost modes, Yukawa-like static potentials, power-counting renormalizability and loss of unitarity should be addressed. In what follows, we review some of those properties, calculate the associated propagator of the theory and discuss the aforementioned issues.

\subsection{Propagator and unitarity for the non-commutative gravity}

As we have seem, applying the prescription given by Eq. (\ref{j10}) after or before linearization yields the same result. Therefore, if we  start directly from the wave operator for the ordinary Einstein-Hilbert action and apply the same procedure, we find the same operator for the (3+1)D higher-order gravity theory. 
It is convenient to work with the so called Barnes-Rives operators in the space of symmetric rank-two tensors \cite{VanNieuwenhuizen:1973fi,Antoniadis:1986tu}
\begin{eqnarray}
\nonumber P^{(2)}_{\mu\nu,\kappa\lambda}&=&\frac{1}{2}\left(\theta_{\mu\kappa}\theta_{\nu\lambda}+\theta_{\mu\lambda}\theta_{\nu\kappa}\right)-\frac{1}{3}\theta_{\mu\nu}\theta_{\kappa\lambda},\label{p2}\\
\nonumber P^{(1)}_{\mu\nu,\kappa\lambda } &=&\frac{1}{2}(\theta_{\mu\kappa}\omega_{\nu\lambda}+\theta_{\mu\lambda}\omega_{\nu\kappa}+\theta_{\nu\lambda}\omega_{\mu\kappa}+\theta_{\nu\kappa}\nonumber \omega_{\mu\lambda}),\label{p1}\\ \nonumber 
P^{(0-s)}_{\mu\nu,\kappa\lambda}&=&\frac{1}{3}\theta_{\mu\nu}\theta_{\kappa\lambda},\label{ps}\\ \nonumber 
P^{(0-w)}_{\mu\nu,\kappa\lambda}&=&\omega_{\mu\nu}\omega_{\kappa\lambda},\label{pw}\\ \nonumber 
P^{(0-sw)}_{\mu\nu,\kappa\lambda}&=&\frac{1}{\sqrt{3}}\theta_{\mu\nu}\omega_{\kappa\lambda},\label{psw}\\ \nonumber 
P^{(0-ws)}_{\mu\nu,\kappa\lambda}&=&\frac{1}{\sqrt{3}}\omega_{\mu\nu}\theta_{\kappa\lambda}\label{pws},\nonumber 
\end{eqnarray}
where $\theta_{\mu\nu}\equiv\eta_{\mu\nu}-\frac{k_{\mu}k_{\nu}}{k^{2}}$ and $\omega_{\mu\nu}\equiv\frac{k_{\mu}k_{\nu}}{k^{2}}$ are, respectively, the usual transverse and longitudinal projection operators. If we rewrite the Einstein-Hilbert action as
\begin{equation}
\mathcal{L}_{R}=\frac{2R}{\kappa^2}\sqrt{-g}=\frac{1}{2}h_{\mu\nu}\mathcal{O}_{R}^{\mu\nu\alpha\beta}h_{\alpha\beta},
\end{equation}
in momentum space we have
\begin{eqnarray}
\nonumber \mathcal{O}^{\mu\nu\alpha\beta}_{R}&=&-\frac{1}{2}\left(\eta^{\mu\alpha}k^{\nu}k^{\beta}+\eta^{\mu\beta}k^
{\nu}k^{\alpha}+\eta^{\nu\beta}k^{\mu}k^{\alpha}\right)+\eta^{\mu\nu}k^{\alpha}k^{\beta}-\eta^{\alpha\beta}k^{\mu}k^{\nu}
+\\ \nonumber &-&\eta^{\mu\nu}\eta^{\alpha\beta}k^2+\frac{1}{2}\left(\eta^{\mu\alpha}\eta^{\nu\beta}+\eta^{\nu\alpha\mu\beta}
\right)k^2\\
&=&k^2\left[P^{(2)}-2P^{(0-s)}\right]^{\mu\nu\alpha\beta}.
\end{eqnarray}
Making the substitution $k^{\alpha}\rightarrow k^{\alpha}(1-k^2\beta\hbar^2)$, to first order in $\beta$ we promptly find
\begin{equation}
\mathcal{O}^{\mu\nu\alpha\beta}_{R(NC)}=k^2(1-2k^2\beta\hbar)\left[P^{(2)}-2P^{(0-s)}\right]^{\mu\nu\alpha\beta}.
\end{equation}
Adding a gauge fixing term, $\mathcal{L}_{gf}=\frac{1}{2\lambda}(\gamma_{\mu\nu}^{,\nu})^2$ ($\gamma_{\mu\nu}\equiv h_{\mu\nu}-\frac{1}{2}\eta_{\mu\nu}h$), and its associated wave operator $\mathcal{O}_{gf}$, the complete form of $\mathcal{O}_{NC}$ is
\begin{eqnarray}
\nonumber\mathcal{O}_{NC}&=&\mathcal{O}_{R(NC)}+\mathcal{O}_{gf}\\ \nonumber
&=&\frac{k^2}{2\lambda}P^{(1)}+\left(k^2-2k^4\beta\hbar^2\right)P^{(2)}+\frac{k^2}{4\lambda}P^{(0-w)}+\\\nonumber&&+\left[\left(\frac{3}{4\lambda}-2\right)k^2+4k^4\beta\hbar^2\right]P^{(0-s)}+\\ &&-\frac{k^2\sqrt{3}}{4\lambda}\left[P^{(0-sw)}+P^{(0-ws)}\right].
\end{eqnarray}
Upon the inversion of $\mathcal{O}_{NC}$, the propagator of the theory is found to be
\begin{eqnarray}
\nonumber \mathcal{O}_{NC}^{-1}&=&\frac{m_{\beta}^2}{k^2(m_{\beta}^2-k^2)}P^{(2)}+\frac{m_{\beta}^2}{2k^2(k^2-m_{\beta}^2)}P^{(0-s)}+\\ \nonumber &&+\frac{2\lambda}{k^2}P^{(1)}+\left[\frac{4\lambda}{k^2}+\frac{3m_{\beta}^2}
{2k^2(k^2-m_{\beta}^2)}\right]P^{(0-w)}+\\&&+\frac{\sqrt{3}m_{\beta}^2}{2k^2(k^2-m_{\beta}^2)}\left[P^{(0-ws)}+P^{(0-sw)}\right],\label{prop}
\end{eqnarray}
where we have defined $m_{\beta}^2\equiv\frac{1}{2\beta\hbar^2}$.
As can be readily seem, the term in $P^{(2)}$ is associated to a spin-2. Besides, we have the expected massless spin-2 graviton and another massive excitation (the scalar particle in $P^{(0-s)}$) with mass $m_{\beta}$ as well. Moreover, since $\beta\geq 0$ there are no tachyons in the model.

Now note that this propagator is very similar to the one obtained form the linear approximation of the higher-order gravity, given by the action \cite{Stelle}
\begin{equation}
\mathcal{S}_{HOG}=\int d^4x\sqrt{-g}\left[\frac{2}{\kappa^2}R+\frac{\alpha}{2} R^2 +\frac{\gamma}{2}R^{\mu\nu}R_{\mu\nu}\right]\label{hog},
\end{equation}
whose propagator can be expressed as
\begin{eqnarray}
\nonumber \mathcal{O}^{-1}_{HOG}&=&\frac{m_{2}^2}{k^2(m_{2}^2-k^2)}P^{(2)}+\frac{m_{0}^2}{2k^2(k^2-m_{0}^2)}P^{(0-s)}+\\ \nonumber &&+\frac{2\lambda}{k^2}P^{(1)}+\left[\frac{4\lambda}{k^2}+\frac{3m_{0}^2}
{2k^2(k^2-m_{0}^2)}\right]P^{(0-w)}+\\&&+\frac{\sqrt{3}m_{0}^2}{2k^2(k^2-m_{0}^2)}\left[P^{(0-ws)}+P^{(0-sw)}\right], \label{prop1}
\end{eqnarray}
where $m^2_2=-\frac{4}{\gamma\kappa^2}$ and $m_0^2=\frac{2}{(3\alpha+\gamma)\kappa^2}$. 

It is immediate to see that equations (\ref{prop}) and (\ref{prop1}) provide the same particle content if we request $m_\beta^2=m_0^2=m_2^2$. Therefore, in this sense, the theory obtained from the first order approximation of the generalized Einstein-Hilbert action is equivalent to its counterpart coming from higher-derivative gravity. 
Going further, contracting the propagator (\ref{prop}) with conserved currents $T^{\mu\nu}(k)$, ($k_{\mu}T^{\mu\nu}=k_{\nu}T^{\mu\nu}=0$), yields
\begin{eqnarray} 
\mathrm{SP}&=&T^{\mu\nu}\left(\mathcal{O}^{-1}_{\mathrm{NC}}\right)_{\mu\nu\alpha\beta}T^{\alpha\beta}\nonumber\\
&=&\left(\frac{1}{k^2}-\frac{1}{k^2-m_\beta^2}\right)\left[T^2_{\mu\nu}-\frac{T^2}{2}\right].
\end{eqnarray}
However, as it is well-known, the tree-level unitarity of a generic model is assured if the residue at each simple pole of ${\mathrm{SP}}$ is $\geq 0$ \cite{accioly3}.
Also, we can  prove that if $m$ is the mass of a generic physical particle related to a given (3+1)D gravitational model and $k$ is the corresponding exchanged momentum, then
\begin{eqnarray}
&\left(T^2_{\mu\nu}-\frac{T^2}{2}\right)&|_{k^2=m^2}>0 \quad\mathrm{ and}\label{res1}\\
&\left(T^2_{\mu\nu}-\frac{T^2}{2}\right)&|_{k^2=0}=0,\label{res2}
\end{eqnarray}
with the additional assumption that $T\geq 0$. 
Therefore, the residues of SP at the poles $k^2=m_\beta^2$ and $k^2=0$ are, using the results of (\ref{res1}) and (\ref{res2}), respectively,
\begin{eqnarray}\label{14}
Res({\mathrm{SP}})\left. \right|_{k^{2}=m_\beta^2}&=&-\left.\left(T^2_{\mu\nu}-\frac{T^2}{2}\right)\right|_{k^{2}=m_\beta^2}<0,\\\label{15}
Res({\mathrm{SP}}) \left.\right|_{k^2=0}&=&\left.(T^2_{\mu\nu}-\frac{T^2}{2})\right|_{k^2=0}=0.
\end{eqnarray}

Thus, this model violates unitarity, but we do not need to worry about it as long as we treat it as an effective field theory until some energy scale given by $m_\beta^2=\frac{1}{2\beta\hbar^2}$, being unitary in this regime.

\subsection{Gravitational bounds on the minimal length}
One way to determine the magnitude of the minimal length is analysing the scale where we expect to find deviations from the usual standard physics. As we have seen, the introduction of a minimal length can lead to higher-derivative models. Moreover, the linearized version of the higher-order gravity theory can be also achieved. In this manner, we could compare the limits on the Newtonian potential in order to set a bound on this minimal length. 
The static gravitational potential obtained from the propagator in (\ref{prop}) for a mass $M$ is 
\begin{equation}
V_{g}=-\frac{GM}{r}\left(1-e^{-r/\sqrt{2\beta_g}}\right).
\end{equation}
Comparing the tests results for the inverse-square law obtained in a Cavendish-like experiment with improved torsion balances yields the following constraint on the minimal length \cite{murata,adelberger},
\begin{equation}
\sqrt{\beta_g}<2.3\times10^{-5}m.
\end{equation}

We could also use existent bounds on the parameters $\alpha$ and $\gamma$ and set an upper limit on the minimal length. The most recent analysis of these parameters, using the deflection of radio waves near the Sun and the gravitational Doppler effect, found $|\gamma|\leq 10^{62}$ and $|\alpha|\leq 10^{78}$ \cite{accioly2,breno}. In this way, we would have $\sqrt{\beta}\approx 10^{-4}m$, one order greater than the one obtained via torsion balances.

It is important to note that, as discussed in Section II, this generalization of the Lorentz algebra manifests itself directly into quantum effects, by means of a modification in the commutation relations. Therefore, we may not be able to find a meaningful result by comparing classical potentials. Moreover, precision tests of the gravitational potential have been carried out only in the range of the Solar System, lacking a more detailed study in smaller (and larger) scales than that. 

Again, one may argue that this whole procedure for find a higher-order theory could have been superseded by usual methods, considering expansions of the curvature or non-minimal couplings in the theory. However, as it will be shown in the next section, the method adopted here have the merit of also yield an higher-order electromagnetic model, parametrized exactly by the same minimal length $\sqrt{\beta}$. Therefore, we can directly link the energy scales of the gravitation and electromagnetism and, even more, take advantage of the precise measurements of the latter to improve the bounds on the former.

\section{The Electromagnetic Case}

Starting from the usual Maxwell's electromagnetic Lagrangian, whose derivatives were replaced by Eq.(\ref{j10}) the Lagrangian reads
\begin{equation}
\mathcal{L}=-\frac{1}{4}\bar{F}_{\mu\nu}\bar{F}^{\mu\nu}, \label{j11}
\end{equation} where $\bar{F}_{\mu\nu}=\nabla_{\mu}A_{\nu}-\nabla_\nu A_\mu=(1+\beta \hbar^2 \Box)(\partial_\mu A_\nu -\partial_\nu A_\mu)=(1+\beta \hbar^2 \Box)F_{\mu\nu}$. After some algebra we have 
\begin{equation} \mathcal{L}=-\frac{1}{4}\Big\{F_{\mu\nu}F^{\mu\nu}+2\beta \hbar^2F_{\mu\nu}\Box F^{\mu\nu}+O(\beta^2\hbar^4)\Big\}, \label{j12}\end{equation} being $F_{\mu\nu}\equiv\partial_{[\mu}A_{\nu]}$, the usual electromagnetic tensor. Hence, neglecting a total derivative term, it follows simply that
\begin{equation}
\mathcal{L}_{eff}=-\frac{1}{4}F_{\mu\nu}F^{\mu\nu}+\frac{\beta\hbar^2}{2}\partial_{\alpha}F_{\mu\nu}\partial^{\alpha}F^{\mu\nu},\label{j13}
\end{equation} or alternatively \cite{gae}
\begin{equation}
\mathcal{L}_{eff}=-\frac{1}{4}F_{\mu\nu}F^{\mu\nu}+\beta \hbar^2(\partial_{\alpha}F^{\beta \alpha})^2.\label{j14}
\end{equation} Therefore, by including the minimal length in the derivative operator we arrive at the higher derivative formulation of the electrodynamics in a rather natural way. We shall interpret this result as follows: as it is well known, higher order terms are good candidates for solving the UV divergence of a given theory. In the case of electromagnetism, the UV divergent sector is due to the divergence upon the charge. Hence, once we introduce a minimal length, the correspondent theory cannot present a divergent behaviour upon the (otherwise dimensionless) charge. 

We note, by passing, that if we had not gave up of the $q\gamma$ term in the derivative operator in eq. (\ref{j9}), the resulting Lagrangian would be 
\begin{eqnarray}
\mathcal{L}_{eff}&=&\left. -\frac{1}{4}\Big\{ F_{\mu\nu}F^{\mu\nu}+2\beta \hbar^2F_{\mu\nu}\Box F^{\mu\nu}-q\gamma \Big(2x_{[\mu}A_{\nu]}F^{\mu\nu}\right.\nonumber\\&+&\left.2\beta\hbar^2 (\Box F_{\mu\nu})x^{[\mu}A^{\nu]}-q\gamma x_{[\mu}A_{\nu]}x^{[\mu}A^{\nu]}\Big)\Big\},\label{j15}\right.
\end{eqnarray} 
jeopardizing the gauge invariance of the resulting theory. 

A similar analysis of the propagator, as it was done in the last section, obtained from eq. (\ref{j13}) shows that unitarity is preserved until a energy scale given by $m_g^2=\frac{1}{\beta\hbar^2}$, where $m_g$ is the mass of the gauge boson that mediates the interaction.

Having established the higher-derivative electromagnetic theory arising from the non-commutative considerations in Section II, we now proceed to find bounds on the minimal length comparing the deviations from the Coulomb's law found in classical experiments, and deviations from the anomalous magnetic moment of the electron, in the QED. Here, just like in the gravitational case, we have found a similar higher-order theory as the one studied by Podolsky or Lee and Wick, whose properties are well-known \cite{podolsky,lee-wick}. In what follows we shall briefly discuss the static potential and the experimental bounds on the parameter $\sqrt{\beta}$.

\subsection{Electromagnetic Bounds on the Minimal Length}
The modified static potential for a charge $Q$ in the higher-derivative electromagnetism given by Lagrangian in equation (\ref{j13}) is

\begin{equation}
V_{e}=\frac{Q}{4\pi r}\left(1-e^{-r/\sqrt{\beta_e}}\right).
\end{equation}

Analysing the deviations from the Coulomb's law obtained from the experiment carried out by Plimpton and Lawton \cite{plimpton,williams}, Accioly and Scatena found an upper limit for the coupling constant of the higher-derivative electromagnetism \cite{accioly}, which yields for the minimal length 

\begin{equation}
\sqrt{\beta_e}< 3.2\times 10^{-10}m.
\end{equation}
This classical result for the electromagnetic case is much lower than the one obtained classically for the gravitational one. We could conjecture, however, that a bound from a quantum theory of gravitation would ensure a better limit, but the lack of such a theory does not enable us to find it.

Fortunately, that is not the case for the electromagnetic sector. In fact, its quantum version, quantum electrodynamics (QED) is one of the most successful and tested theories we have at our disposal.

In this vein, we can constraint the $\beta$ value by considering the measurements of the anomalous magnetic moment of the electron. Comparing the expected value in the higher-derivative version of QED with the deviations of the standard QED from the measurements, we can infer a limit on $\beta$ \cite{accioly,moayedi2}. The experimental and theoretical value predicted by ordinary QED for the anomalous magnetic moment of the electron agree in $1$ part in $10^{10}$, and the corresponding limit on $\beta$ considering the higher-derivative theory is found to be

\begin{equation}
\sqrt{\beta_{qed}}<4.7\times10^{-18} m.\label{bqed}
\end{equation}

It is worth to say that this bound on $\sqrt{\beta_{qed}}$ was obtained considering only contributions arising from QED to the anomalous magnetic moment of the electron, while further contributions appear from electroweak and QCD effects \cite{marquard}. Therefore, a complete analysis on this limit should also include higher-order contributions to these other sectors as well, and a non-Abelian-non-commutative theory should be developed to this end.

\section{Final Remarks}

The Table I summarizes the possible values obtained for $\sqrt{\beta}$. As we can see, the two gravitational bounds, obtained using two distinct approaches (static potential and light bending) afford compatible results, while the comparison with the quantum bound suggests that we must go on higher energies to probe the minimal length.

\begin{table}[h]
\centering
\label{table}
\begin{tabular}{@{}lcccc@{}}
\hline
 & \multicolumn{2}{c}{Gravitation} & \multicolumn{2}{c}{Electromagnetism} \\ \hline
 & \begin{tabular}[c]{@{}c@{}}Static \\ Potential\end{tabular} & \begin{tabular}[c]{@{}c@{}}Light \\ Bending\end{tabular} & \begin{tabular}[c]{@{}c@{}}Static \\ Potential\end{tabular} & \begin{tabular}[c]{@{}c@{}}Anomalous \\ Mag. Moment\end{tabular} \\ \hline 
 & \multicolumn{1}{l}{} & \multicolumn{1}{l}{} & \multicolumn{1}{l}{} & \multicolumn{1}{l}{} \\
\multicolumn{1}{c}{$\sqrt{\beta}$ (m)} & $2.3\times 10^{-5}$  & $5.6\times 10^{-4}$  & $5.1\times10^{-10}$  & $4.7 \times 10^{-18}$  \\  \hline
\end{tabular}
\caption{Bounds on $\sqrt{\beta}$ from different experiments.}
\end{table}

One of the most interesting features of introducing higher-order theories from the minimal length is the fact that both gravity and electromagnetism are linked by the same scale. Since the modifications that we expect in both electromagnetic and gravitational sectors are weighted by the same parameter $\beta$ (e.g., $\beta_{qed}=\beta_g$), we can set the same bound on the minimal length for both theories considering only the QED limit (while we keep longing to come up with a quantum gravitational test in a foreseeable future). 

Conversely, this connection between the minimal length obtained from QED calculations and gravitation allows us to, contrary to what was previously made, set more stringent limits on the parameters $\gamma$ and $\alpha$ of the higher derivative gravity. From $m_\beta=m_2=m_0$,  we found $|\gamma|\approx2\alpha\lesssim 10^{33}$, lowering those limits by practically half of the order of magnitude of the existing bounds \cite{accioly2,breno}.

Summarizing, we have seen that higher derivative models may arise as a consequence of particular scenarios regarding non-commutative spaces by means of a generalized Lorentz algebra endowed with a minimal length. After studying the paradignatic linearized higher order gravity case we were able to explore the electromagnetic case. 

It is clear that both classical gravitational bounds on $\sqrt{\beta}$ are far distant from the quantum bound obtained via QED experiments, or even tests of the Coulomb's law. Yet, those classical limits lie inside the maximum energy allowed to keep the theory unitary. As a matter of fact, the corrections to the gravitational potential including truly quantum effects was calculated by Donoghue \cite{donoghue},
\begin{equation*}
V(r)=-\frac{GM_1M_2}{r}\left[1-\frac{G(M_1+M_2)}{rc^2}-\frac{127}{30\pi^2}\frac{G\hbar}{r^2c^3}\right]
\end{equation*}
and, therefore, the quantum correction is $\frac{G\hbar}{r^2c^3}\approx 10^{-38}$ at $r=1$ fm, what is way too small for our actual experimental capability. Even so, if we expect that effects of a minimal length gives origin to both electromagnetic and gravitational higher-derivatives, we can still get some helpful insights without direct quantum gravity experiments. From the theoretical point of view, it was recently shown that one can obtain similar relations from a non-anticommutative deformation, leading to a generalized uncertainty principle, on a supersymmetric scheme. \cite{faizal}

\section*{Acknowledgements}
The authors would like to express their gratitude to professor A. Accioly, whose insights permeate this work. ES thanks to the CAPES/PNPD grant. JMHS thanks to CNPq for partial support.

\end{document}